\documentclass[conference]{IEEEtran}

\IEEEoverridecommandlockouts

\usepackage{graphicx}
\usepackage{amsmath}
\usepackage{amssymb}
\usepackage{color}
\usepackage{bm}

\usepackage{algorithm}
\usepackage{algpseudocode}
\algnewcommand\algorithmicinput{\textbf{Input:}}
\algnewcommand\INPUT{\item[\algorithmicinput]}
\algnewcommand\algorithmicoutput{\textbf{Output:}}
\algnewcommand\OUTPUT{\item[\algorithmicoutput]}

\usepackage{mathtools}

\newcommand{\Fig}[1]{Fig.~\textup{\ref{#1}}}
\graphicspath{{figures/}}
\usepackage{subcaption}

\newcommand{\vb}{\bm}
\newcommand{\mb}{\bm}

\newcommand{\conj}[1]{\operatorname{conj}(#1)}

\renewcommand{\epsilon}{\varepsiolon}

\def\BibTeX{{\rm B\kern-.05em{\sc i\kern-.025em b}\kern-.08em
    T\kern-.1667em\lower.7ex\hbox{E}\kern-.125emX}}
\usepackage[left=0.625in,right=0.625in,
    top=0.75in,bottom=1.0in,bindingoffset=0cm]{geometry}
\begin{document}

\title{ Data-Aided LS Channel Estimation \\
in Massive MIMO Turbo-Receiver \thanks{The research was carried out at Skoltech and supported by the Russian Science Foundation (project no. 18-19-00673).}}

\author{
\IEEEauthorblockN{Alexander Osinsky, Andrey Ivanov, Dmitry Lakontsev, Roman Bychkov, Dmitry Yarotsky}
\IEEEauthorblockA{\small Skolkovo Institute of Science and Technology\\
    Moscow, Russia
}
  { Alexander.Osinsky@skoltech.ru, an.ivanov@skoltech.ru, d.lakontsev@skoltech.ru, R.Bychkov@skoltech.ru, D.Yarotsky@skoltech.ru }  
  {} 
}


%


\maketitle

\begin{abstract}

In this paper, we propose a new algorithm of iterative least squared (LS) channel estimation for 64 antennas Massive Multiple Input, Multiple Output (MIMO) turbo-receiver. The algorithm employs log-likelihood ratios (LLR) of low-density parity-check (LDPC) decoder and minimum mean square error (MMSE) estimator to achieve soft data symbols. These soft data symbols are further MMSE-weighted again and combined with pilot symbols to achieve a modified LS channel estimate. The modified LS estimate is employed by the same channel estimation unit to enhance turbo-receiver performance via channel re-estimation, as a result, the proposed approach has low complexity and fits any channel estimation solution, which is quite valuable in practice. We analyze both hard and soft algorithm versions and present simulation results of 5G turbo-receiver in the 3D-UMa model of the QuaDRiGa 2.0 channel. Simulation results demonstrate up to 0.3dB performance gain compared to the unweighted hard data symbols utilization in the LS channel re-calculation.

\end{abstract}

 \vskip 0.2cm


\section{Introduction}

Massive Multiple Input, Multiple Output (MIMO) is a key technology of the $5G$ generation wireless communication systems \cite{A1, A1A}. The main feature of the Massive MIMO technology is the use of a large number of antennas in the receiver ($64$, $256$ or more, while in the usual $4$G MIMO there are no more than $8$ antennas) as described in \cite{A2, A1A}. As a result, the $5G$ standard provides opportunities for multiple spectrum reuse in multi-user (MU-MIMO) mode and, therefore, high spectrum efficiency due to joint non-linear detection of users and better interference suppression. Therefore, Massive MIMO technology is the basis of the $5G$ standard Enhanced Mobile Broadband (eMBB) as shown in \cite{A1, A1A}. It can be expected that the number of antennas will increase in the next generations of mobile communications.
However, with a growing number of antennas the channel estimation (CE) accuracy gets worse because of decreasing signal-to-noise (SNR) ratio per antenna. Therefore, CE plays a key role in Massive MIMO efficiency. Finally, there is about $1$dB...$2$dB theoretically proven performance loss compared to an ideal channel estimation, which is much more than CE losses of $0.1$dB...$0.5$dB in $4$G technology as shown in \cite{A3}, \cite{A55} and \cite{A5}.

Turbo CE algorithms can improve estimation accuracy by jointly utilizing pilot symbols and a soft estimate of transmitted data symbols \cite{A4}. The turbo receiver is based on the principle of using decoded data to improve channel estimation. The standard $5G$ channel estimate employs pilot symbols for channel estimation, and after decoding the data, more symbols appear \cite{A2}, represented by the soft decision of the low-density parity-check (LDPC) decoder, therefore, the channel re-estimation will be much more accurate. Moreover, it is used both to improve CE accuracy and MIMO detector quality in the MU-MIMO mode, i.e. when detecting a large number of users \cite{A14}. The turbo receiver consists of iteratively calculating the channel estimate, MIMO detection and LDPC decoding, realizing the so-called “external iteration” in order to improve receiver performance \cite{A6}.

The problem of the most efficient utilization of these data symbols remains open since it includes taking into account a large number of parameters and non-linear relationships between them \cite{A7}. In the CE, there is an acute question of new pilots recounting according to the log-likelihood ratios (LLR) obtained after decoding. This calculation is done through non-linear functions (logarithm, product) and contains threshold processing; moreover, it is necessary to make a joint weighted estimate for new and old pilots, and the calculation of weights depends on many parameters. The straightforward way is to treat all the LDPC-decoded data symbols the same way as pilots, which is called hard turbo decoding. Though it seems intuitively obvious that one should treat pilots better than the decoded data and use a higher coefficient for them and that the coefficient for the user symbols should decrease when the LLR values show large error probability. It was shown in \cite{A4, A6, A7, A8} that such a method significantly improves performance when using the weighted sum of the first soft decoding, while problems of probabilities conversion into complex amplitudes and weighting coefficients calculation are not solved, and, as a rule, are performed empirically. Moreover, turbo equalization causes an extra delay in the received signal processing. The extra delay could result in messages lost and a repeat request occurs. Therefore, both performance and complexity of the data-aided least squared (LS) channel estimation make a significant sense \cite{A8, A8A}. Therefore, there are several existing ways to account for this information, but they lack for more efficiency and less complexity.

\begin{figure}[h]
\centering
\includegraphics[width=1.0\columnwidth]{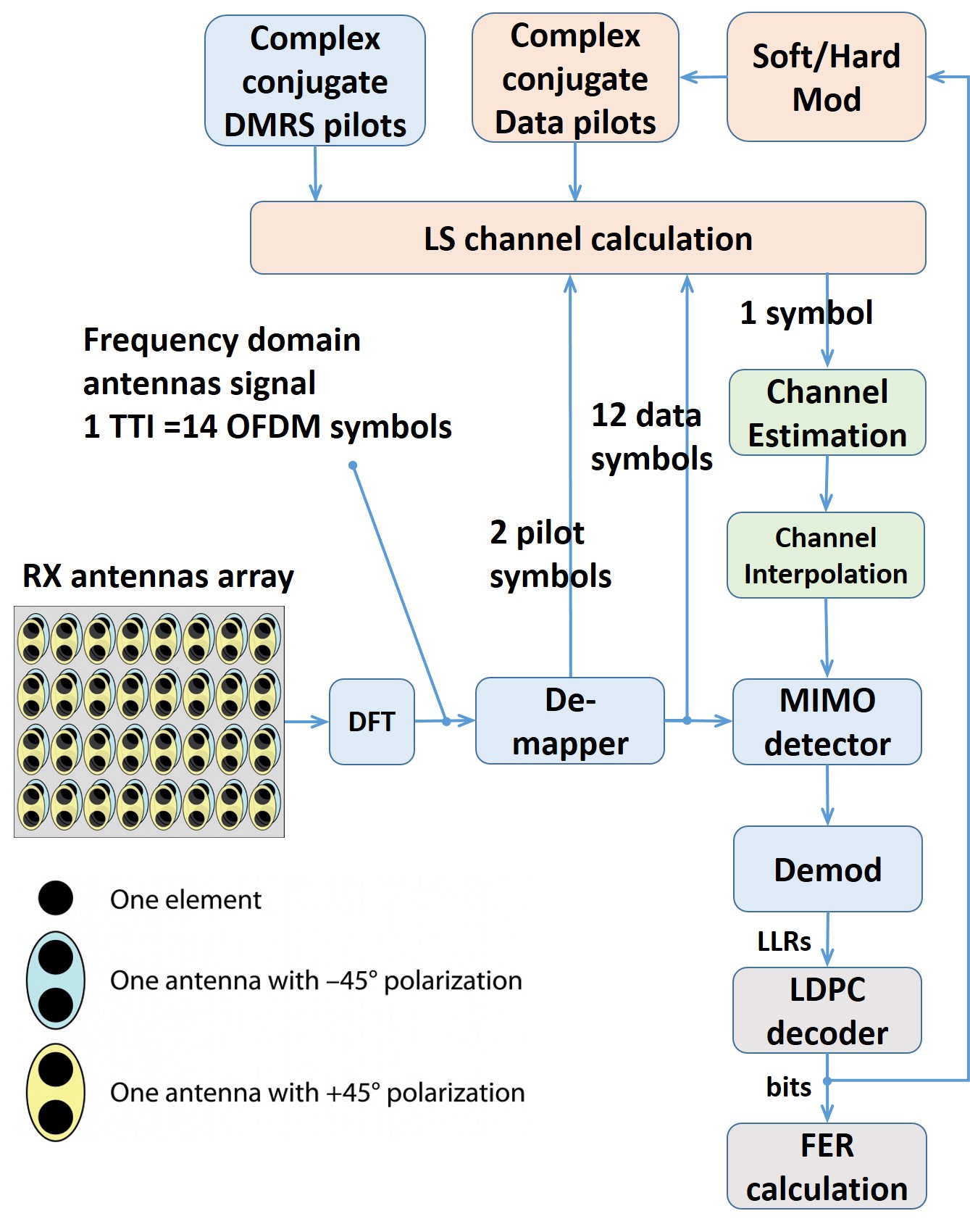}
\caption{
Turbo (iterative) channel estimation
}
\label{fig1}
\end{figure}

In this paper, we construct an efficient soft mapper and a new data-aided LS channel estimation, which employs the information about LLR, i.e. our algorithm re-calculates input signal for the standard CE unit as shown in \Fig{fig1}. The units we are working on are highlighted by red. 

It should be noted that the proposed LS estimation technique is robust to the self CE unit, i.e. our solution can be used with any CE algorithm. We provide both theory and simulations with $5G$ turbo-receiver in Quadriga channel \cite{A9}. It can help to understand which techniques are useful in constructing soft turbo LS estimations and which approaches are doomed to fail.

\section{Simulation tool}

\begin{figure}[h]
\centering
\includegraphics[width=1.0\columnwidth]{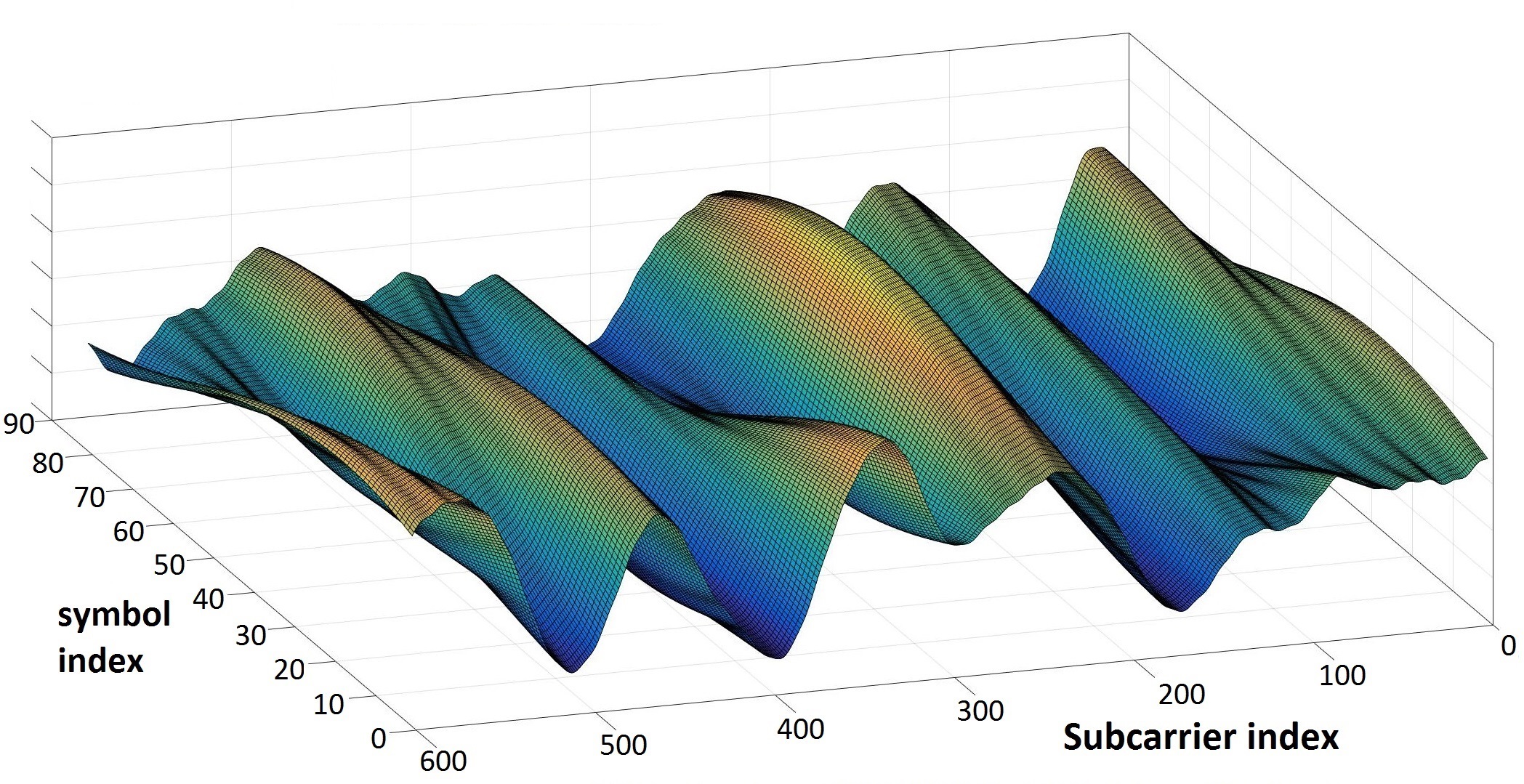}
\caption{
Magnitude spectrum of QuaDRiGa channel
}
\label{fig2}
\end{figure}

The MMSE algorithm from \cite{A13, A12} and \cite{A122} performs the MIMO detector function in \Fig{fig1} and antenna array consists of $64$ antennas ($2$ co-located sub-arrays of size $8 \times 4$ with different polarization). We utilize (144,288) LDPC code with the Min-Sum decoding algorithm in the receiver end \cite{A10,A11} . The DFT-based channel estimation was implemented as described in \cite{A5}. QuaDRiGa, short for "QUAsi Deterministic RadIo channel GenerAtor" \cite{A9}, was used to generate realistic radio channel responses in system-level simulations of $5G$ scenarios. We test our algorithms with $64$ antennas MIMO in 3D-UMa non-line of sight (NLOS) scenarios for single antenna users with a speed of $5$ km/h. Our simulation set consists of $140$ propagation scenarios with $32$ additive white noise seeds per each scenario. A short $3D$ fragment of the channel magnitude spectrum is shown in \Fig{fig2}.

\section{Soft modulator}\label{llr-sec}

To construct a soft (or hard) modulator shown in \Fig{fig1} we first of all need to determine probabilities of attaining each QAM point. These probabilities are computed as follows. Each lattice node has a binary code $c \in \left\{ 0, 1 \right\}^Q$, where $Q$ is the modulation order. Each bit has a corresponding LLR value $LLR_k$, which defines the probability of it to be equal to $1$, $k = \overline{1, Q}$. The LLR value is given by:
\[
  LLR_k = \ln \frac{p_k(0)}{p_k(1)},\\
\]
where $p_k(0)$ and $p_k(1)=p_k$ are the probability of bits "0" and "1" receiving respectively. Since $p_k(0)+p_k(1)=1$, the LLR equation is given by: 
\[
  LLR_k = \ln \frac{1-p_k}{p_k},
\]
Therefore, the probabilities can be computed as:
\[
  p_k = \frac{1}{1 + e^{LLR_k}}
\]
Total probability for each lattice node (QAM point) can be calculated as a product of all $p_k$ for the bits equal to 1 and $1 - p_k$ for all the bits equal to $0$. Let us denote these probabilities by $P_j$, $j = \overline{1, 2^Q}$.

Let $\vb{q} \in \mathbb{C}^{2^Q}$ be the vector of all QAM points. For example, for QPSK we have $q = \left[ {\begin{array}{*{20}{c}}
   {\tfrac{{1 + i}}{ \sqrt{2}}} & {\tfrac{{1 - i}}{\sqrt{2}}} & {\tfrac{{ - 1 + i}}{\sqrt{2}}} & {\tfrac{{ - 1 - i}}{\sqrt{2}}}  \\ 
\end{array} } \right]$. Let us denote by $x$ a random variable, which is equal to $q_j$ with probability $P_j$, that is
\[
  \mathcal{P}\left(x = q_j\right) = P_j
\]
Value of $j$, corresponding to the highest probability $P_j$ corresponds to the "hard" QAM point (hard modulator). To construct a soft modulator we need to use the expected value.

Expectations can be computed the same way as for any other discrete random variable. For instance,
\begin{equation}\label{expectation}
  \mathbb{E} x = \mathop\sum\limits_{j = 1}^{2^Q} \mathcal{P}\left(x = q_j\right) q_j = \mathop\sum\limits_{j = 1}^{2^Q} P_j q_j
\end{equation}

Later we will use vector notation. Equation (\ref{expectation}) then shows how to compute expectations for each element $x$ of any data vector $\vb{x}$.

\section{Data-aided LS channel estimation}

Let $\mb{Y} \in \mathbb{C}^{N_{ofdm} \times N_{used} \times N_{rx}}$ be the received data, where $N_{ofdm}$ is the total number of orthogonal frequency division multiplexing (OFDM) symbols in one time transmission interval (TTI) : 
\[
N_{ofdm} = N_{data} + N_{pilot},
\]
where $N_{data}=12$ is the number of data symbols, $N_{pilot}=2$ is the number of pilot symbols, $N_{used}=RB_{num} \times RB_{size}$ is the total number of subcarriers, $1$ resource block (RB) contains $RB_{size}=12$ subcarriers, $RB_{num}$ is the number or reserved RB, $N_{rx}$ is the total number of antennas. Let $\mb{H} \in \mathbb{C}^{N_{used} \times N_{rx}}$ be the channel matrix. Let us denote by $\vb{y} \in \mathbb{C}^{N_{ofdm}}$ a single vector of data from $\mb{Y}$, corresponding to the element $h$ from $\mb{H}$. It can be represented as:
\[
  \vb{y} = h \vb{x} + \vb{e}
\]
or elementwise
\[
  y_i = h x_i + e_i,
\]
where $x_i$ is the lattice node used to send the data and $e_i$ is an additive white gaussian noise. For pilots the equation is:
\[
  y_i = h p + e_i,
\]
where $p^2$ is the power ratio of the pilot signal compared to data symbols (usually, pilot symbols are transmitted with $2$ times higher power compared to data ones). 

Hereafter we will treat data symbols the same way as pilot symbols, but with higher uncertainty. For example, if there are two pilots, we can set $x_1 = p$ and $x_2 = p$ and other elements of $\vb{x}$ will correspond to the data. In general, $x_i$ are random variables. For pilot symbols the $p$ value is known with $100\%$ certainty, while data symbols are given as discrete randomly distributed complex variables, which take values on the lattice. Probabilities of attaining each value can be determined using the knowledge of $LLRs$ (see section \ref{llr-sec}).

In case when only pilots are used to decode the data or all $x_i$ are rescaled to $1$ and the speed of the user is not high, one can use just a simple averaging over pilot symbols before passing data to the MIMO detector again (non-turbo version, where the input of Channel Estimation unit in \Fig{fig2} is calculated from $2$ pilot symbols). The same thing can, of course, be done in the turbo mode if we use a hard version by fixing $x_i$. However, there should be a better way.

Our task is to estimate element $h$ of the matrix $\mb{H}$, corresponding to data vector $\vb{y}$. We will achieve this by constructing the linear minimum mean square error (MMSE) estimate. The general MMSE estimate $\vb{\hat u}$ of an unknown random variable $\vb{u}$ based on the observation $\vb{z}$ can be expressed as:
\[
  \vb{\hat u} = \mathbb{E} \left( \vb{u} \vb{z^H} \right) \left( \mathbb{E} \left( \vb{z} \vb{z^H} \right) \right)^{-1} \vb{z}
\]
In our case $\vb{u} = h$ is a single value and $\vb{z} = \vb{y}$ is a vector, therefore, the averaged product
\[
  \mathbb{E} \left( h \vb{y^H} \right) = \mathbb{E} \vb{x^H} \mathbb{E} | h |^2 \in \mathbb{C}^{N_{ofdm}}
\]
represents a row vector and the correlation matrix for $\vb{y}$ can be calculated as:
\[
  \mathbb{E} \left( \vb{y} \vb{y^H} \right) = \mathbb{E} \left( \vb{x} \vb{x^H} \right) \mathbb{E} | h |^2 + \mathbb{E} | e_i |^2 \mb{I} \in \mathbb{C}^{N_{ofdm} \times N_{ofdm}}
\]
Therefore, we get for the MMSE estimate $\hat h$ as:
\begin{equation}\label{soft-gen}
  \hat h = \mathbb{E} \vb{x^H} \left( \mathbb{E} \left( \vb{x} \vb{x^H} \right) + \sigma^2 \mb{I} \right)^{-1} y,
\end{equation}
where $\sigma^2 = \mathbb{E} | e_i |^2 / \mathbb{E} | h |^2$. Equation (\ref{soft-gen}) is what we are going to use to define the soft modulator in \Fig{fig1}. We don't use it directly, because it involves matrix inverse. Let us simplify it by explicitly calculating the inverse of the correlation matrix.

First of all, we note that non-diagonal terms of the correlation matrix $\mathbb{E} \left( \vb{x} \vb{x^H} \right)$ are equal to non-diagonal terms of $\mathbb{E} \vb{x} \mathbb{E} \vb{x^H}$. Indeed, non-diagonal values are equal to $\mathbb{E} \left( x_i \conj{x_j} \right)$ with $i \ne j$. QAM points from different data symbols are independent, therefore, for $i \ne j$ we have $\mathbb{E} \left( x_i \conj{x_j} \right) = \mathbb{E} x_i \mathbb{E} \conj{x_j}$. For diagonal elements ($i = j$) we have the equality:
\[
  \mathbb{E} \left( x_i \conj{x_i} \right) = \mathbb{E} x_i \mathbb{E} \conj{x_i} + \left( \mathbb{E} |x_i|^2 - \left| \mathbb{E} x_i \right|^2 \right)
\]
Therefore,
\[
  \mathbb{E} \left( \vb{x} \vb{x^H} \right) = \mathbb{E} \vb{x} \mathbb{E} \vb{x^H} + diag \left( \mathbb{E} |x_i|^2 - \left| \mathbb{E} x_i \right|^2 \right)
\]
and
\[
  \mathbb{E} \left( \vb{x} \vb{x^H} \right) + \sigma^2 \mb{I} = \left(\sigma^2 \mb{I} + diag \left( \mathbb{E} |x_i|^2 - \left| \mathbb{E} x_i \right|^2 \right) \right) + \mathbb{E} \vb{x} \mathbb{E} \vb{x^H}
\]
The first term is a diagonal matrix while the second term is a rank one matrix. Inverse of such combination can be calculated explicitly using rank one update formula. Let us denote:
\[
  \mb{A} = \sigma^2 \mb{I} + diag \left( \mathbb{E} |x_i|^2 - \left| \mathbb{E} x_i \right|^2 \right)
\]
and
\[ 
  \vb{\phi} = \mathbb{E} \vb{x},
\]
Then the r.h.s. of equation (\ref{soft-gen}) can be calculated as:
\[
\begin{gathered}
  \mathbb{E} \vb{x^H} \left( \left( \sigma^2 \mb{I} + diag \left( \mathbb{E} |x_i|^2 - \left| \mathbb{E} x_i \right|^2 \right) \right) + \mathbb{E} \vb{x} \mathbb{E} \vb{x^H} \right)^{-1} \hfill \\
  = \vb{\phi^H} \left( \mb{A} + \vb{\phi} \vb{\phi^H} \right)^{-1} = \vb{\phi^H} \left(\mb{A}^{-1} - \frac{\mb{A}^{-1} \vb{\phi} \vb{\phi^H} \mb{A}^{-1}}{1 + \vb{\phi^H} \mb{A}^{-1} \vb{\phi}} \right) \hfill \\
  = \frac{\vb{\phi^H} \mb{A}^{-1}}{1 + \vb{\phi^H} \mb{A}^{-1} \vb{\phi}} \hfill
\end{gathered}
\]
After substitution we can rewrite the result in elementwise notation as follows:
\begin{equation}\label{soft-final}
  \hat h = \mathop\sum\limits_{i = 1}^{N_{ofdm}} \frac{\mathbb{E} \left( \conj{x_i} y_i \right)}{\mathbb{E} |x_i|^2 + \sigma^2 + \mathop\sum\limits_{j \neq i} \left| \mathbb{E} x_j \right|^2 \frac{\mathbb{E} |x_i|^2 - \left| \mathbb{E} x_i \right|^2 + \sigma^2}{\mathbb{E} |x_j|^2 - \left| \mathbb{E} x_j \right|^2 + \sigma^2}}
\end{equation}

To use equation (\ref{soft-final}) one needs to find expectations for $x_i$ and $| x_i |^2$ based on LLR and substitute them to get the appropriate scaling of $y_i$. It should be noted, however, that the value of $\sigma^2$ depends on both application scenario and receiver parameters (code rate, code length, MIMO detector algorithm and other). The leftover noise power should be approximately proportional to the original noise power, but much less than it. Therefore, one can find $\sigma^2$ by fitting the corresponding coefficient to minimize the frame error rate (FER).

\section{Complexity reduction}

To further simplify the equation (\ref{soft-final}) let us assume that the average variance after the soft modulator is proportional to the original noise power:
\[
\begin{aligned}
  avg \left( \mathbb{E} |x_i|^2 - \left| \mathbb{E} x_i \right|^2 \right) & \mathop{=}\limits_{}^{def} \frac{\mathop\sum\limits_{i = 1}^{N_{data}} \mathop\sum\limits_{j = 1}^{N_{used}} \left( \mathbb{E} |x_i^j|^2 - \left| \mathbb{E} x_i^j \right|^2 \right)}{N_{used}N_{data}} \\
  & \approx \sigma^2 / C,
\end{aligned}
\]
where $j$ is the subcarrier index. We can approximate $\sigma^2$ as:
\[
  \sigma^2 \approx C \cdot \frac{\mathop\sum\limits_{i = 1}^{N_{data}} \mathop\sum\limits_{j = 1}^{N_{used}} \left( \mathbb{E} |x_i^j|^2 - \left| \mathbb{E} x_i^j \right|^2 \right)}{N_{used}N_{data}}
\]
Constant $C$ should not depend on the noise power and can be fitted. Parameter $C$ was optimized by the genetic algorithm (GA) to minimize turbo-receiver FER. According to our simulations in Quadriga channel, $\sigma^2$ is an order of magnitude higher than the variance of $x_i$, meaning the soft modulator decreases the noise quite sufficiently. Consequently, we can approximate the ratios as follows:
\[
  \frac{\mathbb{E} |x_i|^2 - \left| \mathbb{E} x_i \right|^2 + \sigma^2}{\mathbb{E} |x_j|^2 - \left| \mathbb{E} x_j \right|^2 + \sigma^2} \approx 1
\]
Finally, we achieve the equation:
\begin{equation}\label{soft-simple}
  \hat h \approx \mathop\sum\limits_{i = 1}^{N_{ofdm}} \frac{ \mathbb{E} \left(\conj{x_i} y_i \right)}{\mathbb{E} |x_i|^2 + \mathop\sum\limits_{j \neq i} \left| \mathbb{E} x_j \right|^2 + C \cdot avg \left( \mathbb{E} |x_j|^2 - \left| \mathbb{E} x_j \right|^2 \right)}
\end{equation}
Here and after this algorithm is named {\bf soft param}. Finally, in our experiments, the GA optimization results in $C \approx 17$.

We compare our {\bf soft param} approach (\ref{soft-simple}) with {\bf hard} one and some other intuitive techniques, which are briefly described here. A straightforward {\bf hard} version of the modulator in data-aided LS channel estimation is given by:
\begin{equation}
\hat h = \frac{1}{N_{ofdm}} \mathop\sum\limits_{i = 1}^{N_{ofdm}} \frac{ \conj{[x_i]} y_i}{ \left|[x_i]\right|^2},
\end{equation}
where instead of expectations $\mathbb{E} \conj{x_i}$ we use the values of $x_i$ from the most probable lattice node. Denote these values by $[x_i]$. For pilots $[x_i] = x_i$ just returns the appropriate scale. 

The simplest soft turbo receiver can be obtained by applying generalized least squares to the observation process of:
\[
  y_i = h \mathbb{E} x_i + h \left( x_i - \mathbb{E} x_i \right) + e_i = h \mathbb{E} x_i + e'_i
\]
The unbiased estimate can be calculated as:
\begin{equation}\label{unbiased}
\hat h = \frac{\mathop\sum\limits_{i = 1}^{N_{ofdm}} \mathbb{E} \left( \conj{x_i} y_i \right)}{\mathop\sum\limits_{i = 1}^{N_{ofdm}} \mathbb{E} x_i \mathbb{E} \conj{x_i}},
\end{equation}
Here and after the algorithm (\ref{unbiased}) is named {\bf soft unbiased}. 
However, equation (\ref{unbiased}) completely ignores the noise, while most error comes from additive noise and not from soft modulation mistakes. In practice, quantization approach can also be applied to improve the {\bf soft unbiased} one performance, where instead of expectations $\mathbb{E} \conj{x_i}$ we again use the values of $x_i$ from the most probable lattice node:
\begin{equation}\label{hard_good}
\hat h = \frac{\mathop\sum\limits_{i = 1}^{N_{ofdm}} \conj{[x_i]} y_i}{\mathop\sum\limits_{i = 1}^{N_{ofdm}} \left|[x_i]\right|^2}
\end{equation}
The algorithm (\ref{hard_good}) is named {\bf hard weighted}. The advantage of this approach compared to the simplest {\bf hard} version is that it correctly minimizes the additive noise by using a larger weight for received signals with larger power.

\section{Simulation results}

\begin{figure}[h]
\centering
\includegraphics[width=1.0\columnwidth]{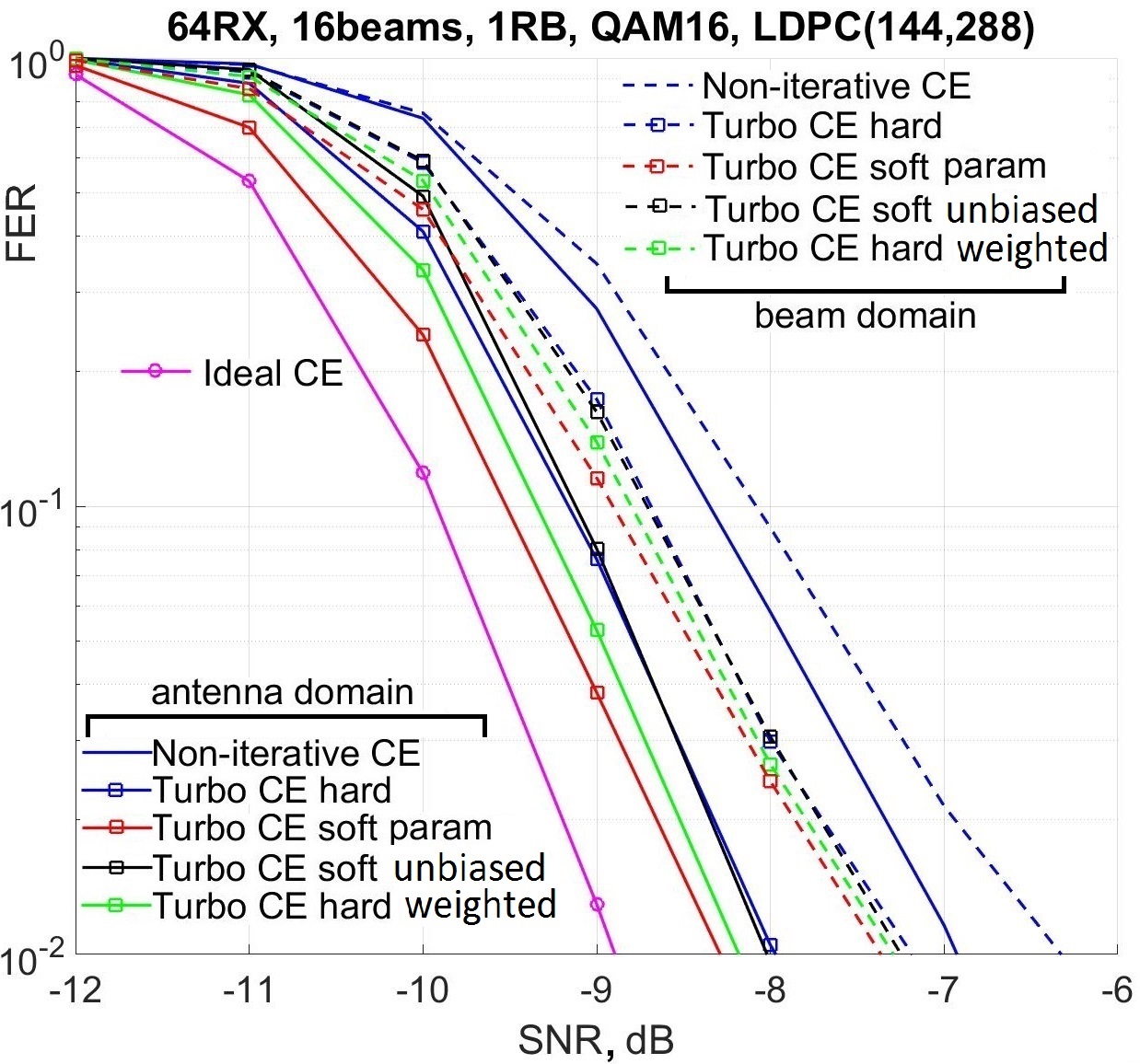}
\caption{
Turbo-receiver performance for 1RB, QAM16
}
\label{fig3}
\end{figure}

\begin{figure}[h]
\centering
\includegraphics[width=1.0\columnwidth]{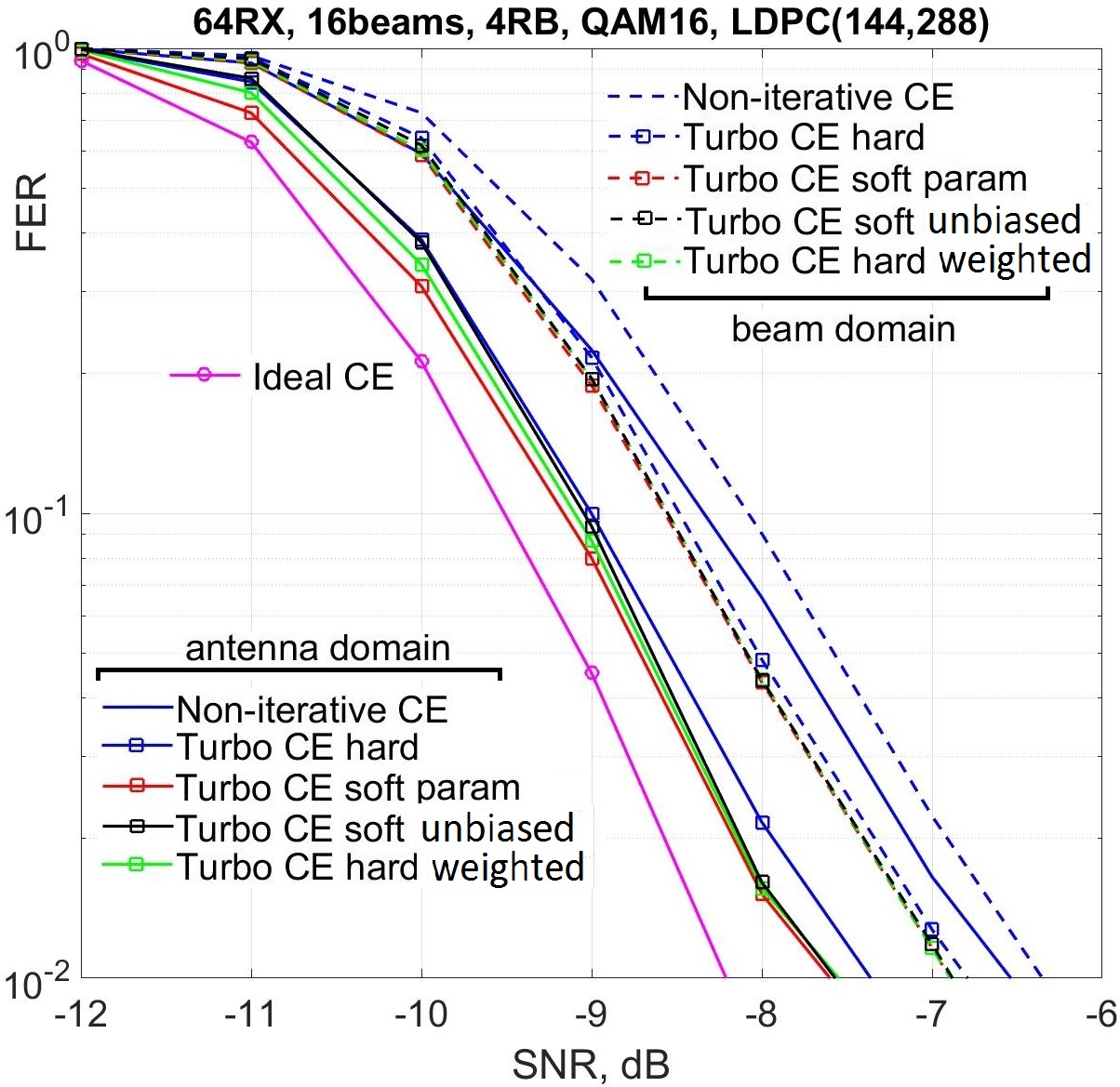}
\caption{
Turbo-receiver performance for 4RB, QAM16
}
\label{fig4}
\end{figure}

\begin{figure}[h]
\centering
\includegraphics[width=1.0\columnwidth]{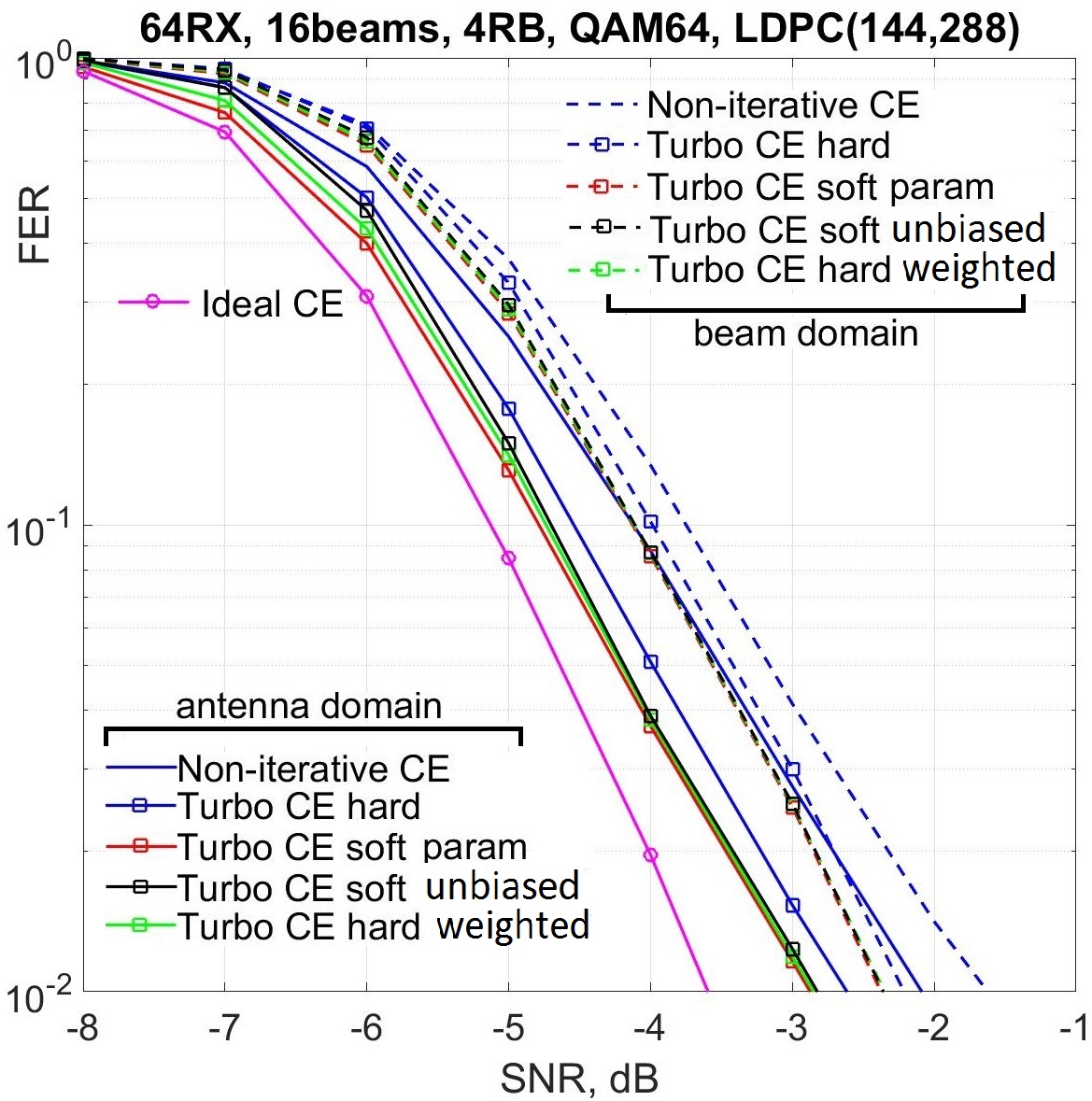}
\caption{
Turbo-receiver performance for 4RB, QAM64
}
\label{fig5}
\end{figure}

One of the methods to reduce complexity is based on CE implementation in a beam domain. The beam domain CE is based on the input signal transformation from antenna to beam domain via multiplying antennas signal by the transform matrix. As a result, signal dimensionality is reduced from $N_{RX}=64$ omnidirectional antennas to, e.g. $N_{PORT}=16$ beams focused on the channel scatterers. Then both channel estimation and MIMO detection algorithms are performed in the beam domain, resulting in less complexity. Usually, the beam transform matrix is calculated from sounding reference signals (SRS), intended to achieve the channel state information (CSI). CSI describes how the signal propagates from the target user and represents the combined effect of scattering, fading, and power delay. Therefore, we test data-aided LS estimations for both antenna and beam domains in $1RB$ and $4RB$ bandwidths in both turbo and non-turbo modes. Simulation results are presented in \Fig{fig3}, \Fig{fig4} and \Fig{fig5} for code rate of $0.5$ and modulation orders of $QAM16$ and $QAM64$.

\section{Acknowledgement}

We would like to thank Prof. Alexey Frolov for his brilliant comments and suggestions. The authors acknowledge the use of Zhores for obtaining the results presented in this paper.

\section{Conclusion}

The proposed soft algorithm of data-aided LS channel estimation demonstrates up to $0.3$dB performance gain in frame error rate compared to the straightforward hard version. This gain is quite valuable for the turbo-receiver, which brings about $0.8$dB gain compared to the non-turbo one. Our method doesn't depend on the implementation of the channel estimation unit and, therefore, can be applied in any receiver software. To achieve an extra performance gain we applied a genetic algorithm to optimize parameter $C$ in a training set of channel realizations. We have validated the trained value of $C=17$ for both antenna and beam domain channel estimates in 3D-UMa NLOS configuration of $5G$ Quadriga channel for $5$km/h user. The algorithm was compared with other data-aided LS channels estimates in different bandwidths and modulation orders to prove the robustness.


\begin{thebibliography}{99}


\bibitem{A1}
5G PPP Architecture Working Group,
\newblock View on 5G Architecture,
\newblock \emph {Version 3.0}, 2019.

\bibitem{A1A}
M. Shafi et al., 
\newblock 5G: A Tutorial Overview of Standards, Trials, Challenges, Deployment, and Practice,
\newblock \emph {IEEE Journal on Selected Areas in Communications}, vol. 35, no. 6, pp. 1201-1221, June 2017.

\bibitem{A2}
https://www.3gpp.org

\bibitem{A3}
H. Xie, F. Gao and S. Jin, 
\newblock An Overview of Low-Rank Channel Estimation for Massive MIMO Systems, 
\newblock \emph {IEEE Access}, vol. 4, pp. 7313-7321, 2016.

\bibitem{A55}
A. Osinsky, A. Ivanov, D. Yarotsky,
\newblock Theoretical Performance Bound of Uplink Channel Estimation Accuracy in Massive MIMO,
\newblock \emph {2020 IEEE International Conference on Acoustics, Speech and Signal Processing (ICASSP)}, Barcelona, 2020.

\bibitem{A5}
H. Al-Salihi, M. R. Nakhai and T. A. Le, 
\newblock DFT-based Channel Estimation Techniques for Massive MIMO Systems,
\newblock \emph {2018 25th International Conference on Telecommunications (ICT)}, St. Malo, 2018, pp. 383-387.

\bibitem{A4}
J. Ma and L. Ping, 
\newblock Data-Aided Channel Estimation in Large Antenna Systems,
\newblock \emph {IEEE Transactions on Signal Processing}, vol. 62, no. 12, pp. 3111-3124, June15, 2014.

\bibitem{A14}
A. Ivanov, A. Savinov and D. Yarotsky,
\newblock Iterative Nonlinear Detection and Decoding in Multi-User Massive MIMO,
\newblock \emph {15th International Wireless Communications and Mobile Computing Conference (IWCMC)}, Tangier, Morocco, 2019, pp. 573-578.

\bibitem{A6}
M. Khalighi, J. J. Boutros and J. Helard, 
\newblock Data-aided channel estimation for turbo-PIC MIMO detectors,
\newblock \emph {IEEE Communications Letters}, vol. 10, no. 5, pp. 350-352, May 2006.

\bibitem{A7}
M. Ju, L. Xu, L. Jin and D. Defeng Huang, 
\newblock Data aided channel estimation for massive MIMO with pilot contamination,
\newblock \emph {2017 IEEE International Conference on Communications (ICC)}, Paris, 2017, pp. 1-6.

\bibitem{A8}
Y. Takano and H. Su, 
\newblock A Low-Complexity LS Turbo Channel Estimation Technique for MU-MIMO Systems,
\newblock \emph {IEEE Signal Processing Letters}, vol. 25, no. 5, pp. 710-714, May 2018.

\bibitem{A8A}
F. Jiang, C. Li, Z. Gong, K. Hao, S. Liu and Y. Zhang, 
\newblock Iterative Approaches for Massive MIMO Uplink Processing Under Imperfect Channel Conditions,
\newblock \emph {IEEE Transactions on Vehicular Technology}, vol. 68, no. 4, pp. 3642-3654, April 2019.

\bibitem{A13}
A. Ivanov, D. Yarotsky, M. Stoliarenko and A. Frolov,
\newblock Smart Sorting in Massive MIMO Detection,
\newblock \emph {14th International Conference on Wireless and Mobile Computing, Networking and Communications (WiMob)}, Limassol, 2018, pp. 1-6.

\bibitem{A12}
A.~Ivanov, S.~Kruglik, D.~Lakontsev,
\newblock Cloud MIMO for Smart Parking System,
\newblock \emph {IEEE 87th Vehicular Technology Conference (VTC-Spring)}, 2018

\bibitem{A122}
A.~Ivanov, A.~Osinsky, D.~Lakontsev, D. Yarotsky,
\newblock High performance interference suppression in multi-user Massive MIMO detector,
\newblock \emph {IEEE 91th Vehicular Technology Conference (VTC-Spring)}, 2020

\bibitem{A9}
http://quadriga-channel-model.de/

\bibitem{A10}
R.~Tanner,
\newblock A recursive approach to low complexity codes.
\newblock \emph{IEEE Trans. Inf. Theory}, 
vol. 27, no. 5, pp. 533--547, Sep. 1981.

\bibitem{A11}
R.~G.~Gallager, 
\newblock \emph{Low-Density Parity-Check Codes}.
\newblock Cambridge: MIT Press, 1963.






\end{thebibliography}
\end{document}